\documentclass[%
 reprint,
 superscriptaddress,
 amsmath,amssymb,
 aps, 
 pra,
]{revtex4-2}

\usepackage{graphicx}
\usepackage{dcolumn}
\usepackage{bm}

\begin{document}

\preprint{APS/123-QED}

\title{Optical excitation and probing of the antiferromagnetic modes with non-uniform in-depth distribution in birefringent antiferromagnetic crystals}

\author{A.A. Voronov}
\email{andrey.a.voronov@gmail.com}
\affiliation{Moscow Institute of Physics and Technology, National Research University, Dolgoprudny, Moscow, 141701 Russia}
\affiliation{Russian Quantum Center, 121205 Moscow, Russia}

\author{D.O. Ignatyeva}
\affiliation{Russian Quantum Center, 121205 Moscow, Russia}
\affiliation{Faculty of Physics, Lomonosov Moscow State University, 119991 Moscow, Russia}
\affiliation{Institute of Physics and Technology, V.I. Vernadsky Crimean Federal University, 295007 Simferopol, Crimea}

\author{A.K. Zvezdin}
\affiliation{Russian Quantum Center, 121205 Moscow, Russia}
\affiliation{Prokhorov General Physics Institute, Russian Academy of Sciences, 119991, Moscow, Russia}

\author{T.B. Shapaeva}
\affiliation{Faculty of Physics, Lomonosov Moscow State University, 119991 Moscow, Russia}

\author{V.I. Belotelov}
\affiliation{Russian Quantum Center, 121205 Moscow, Russia}
\affiliation{Faculty of Physics, Lomonosov Moscow State University, 119991 Moscow, Russia}
\affiliation{Institute of Physics and Technology, V.I. Vernadsky Crimean Federal University, 295007 Simferopol, Crimea}

\date{\today}

\begin{abstract}
Optical pump-probe setups are commonly used for excitation and investigation of the spin dynamics in various types of magnetic materials. However, usually the spatially homogeneous excitation is considered. In the present study we describe an approach for optical excitation of the nonuniform THz spin dynamics and for probing its spatial distribution inside a magnetic crystal. We propose to illuminate a crystal with laser pulses of properly adjusted polarization to benefit from a strong optical birefringence inherent to the crystal. It results in an unusual behavior of the effective magnetic field generated by the pulses due to the inverse Faraday effect and the peculiar sign-changing dependence of the direct Faraday effect inside the crystal. The study is performed exemplary for yttrium orthoferrite crystal although the proposed approach is applicable for various magnetic materials with optical anisotropy. 
\end{abstract}

\maketitle


Studies of spin system dynamics concern different rapidly growing fields of technology such as modern telecommunication technologies ~\cite{neusser2009magnonics,cornelissen2015long}, quantum computing~\cite{ganzhorn2016magnon,lachance2019hybrid}, new approaches for magnetic data recording and reading~\cite{ignatyeva2019plasmonic,shcherbakov2015ultrafast,im2019all}, light modulation at GHz frequencies~\cite{ignatyeva2020all, dadoenkova2017faraday, voronov2020magneto, dadoenkova2016transverse}, biosensing~\cite{borovkova2020high}, and magnetometry~\cite{ignatyeva2021vector}. Antiferromagnetic materials, such as YFeO$_3$, DyFeO$_3$ are among promising candidates for practical use in high-speed devices since they possess great magneto-optical response~\cite{tabor1970visible,kahn1969ultraviolet,chetkin} and allow for the excitation of quasi-antiferromagnetic modes at nearly THz frequencies~\cite{iida2011spectral,bossini2017femtosecond,zhou2012terahertz,yamaguchi2010coherent,kim2017field,bossini2019laser,kim2014coherently,jin2013single,reid2015terahertz,kimel2005ultrafast,kimel2006optical,mikhaylovskiy2015ultrafast}. Although these modes can be excited by the THz electromagnetic pulses~\cite{kim2014coherently,jin2013single,reid2015terahertz}, optical excitation via the femtosecond laser pulse is also very promising since it provides local (at micron and even submicron scales) and tunable impact via various optomagnetic effects, including the inverse Faraday effect (IFE)~\cite{kimel2005ultrafast,savochkin2017generation,kimel2006optical,mikhaylovskiy2015ultrafast,im2017third,de2017effect,chernov2020all,jackl2017magnon}. 

IFE is related to a Raman-like coherent optical scattering process and does not require absorption of light thus providing two main advantages: the optical impact is instantaneous and non-thermal~\cite{kimel2005ultrafast}. In an isotropic magnetic medium the induced via IFE effective magnetic field is maximum for the circularly polarized incident radiation and absent for the linearly polarized one. The IFE induced magnetic field is parallel to the wavevector of light. However, in the optically anisotropic materials the situation is getting more complicated. Due to the conversion of light polarization from initial circular to elliptical and then linear one while propagating through the media the distribution of the IFE field becomes spatially nonuniform. Moreover, generally the orientation of this field may not be parallel to the light wavevector~\cite{volkov2002inverse}. Usually this phenomena is reported to complicate the analysis of the spin dynamics~\cite{iida2011spectral} and several studies were carried out to find out how to interpret the observed probe signal with respect to the pump~\cite{iida2011spectral}, probe~\cite{woodford2007interpreting} or both of the pulse polarizations~\cite{de2017effect}. 

Here we show that, in contrast to the well-known situation in the isotropic materials, optical anisotropy provides a unique tool to excite spatially nonuniform spin dynamics that can not be excited by light in isotropic materials due to the homogeneity of the IFE field inside the pumping area. Moreover, in the isotropic materials the spin modes having some spatial distribution along the sample thickness can not be probed efficiently since the total Faraday rotation of the probe polarization becomes zero if there is an equal amount of spins precessing in opposite phases. Here we show that optical anisotropy makes it possible to probe the temporal dynamics of the nonuniform spin oscillations. Moreover, it provides a novel approach to perform a magneto-optical probe of the spin mode profile spatial distribution inside a magnetic crystal. 

\begin{figure}[h!]
    \centering
    \includegraphics[width=\linewidth]{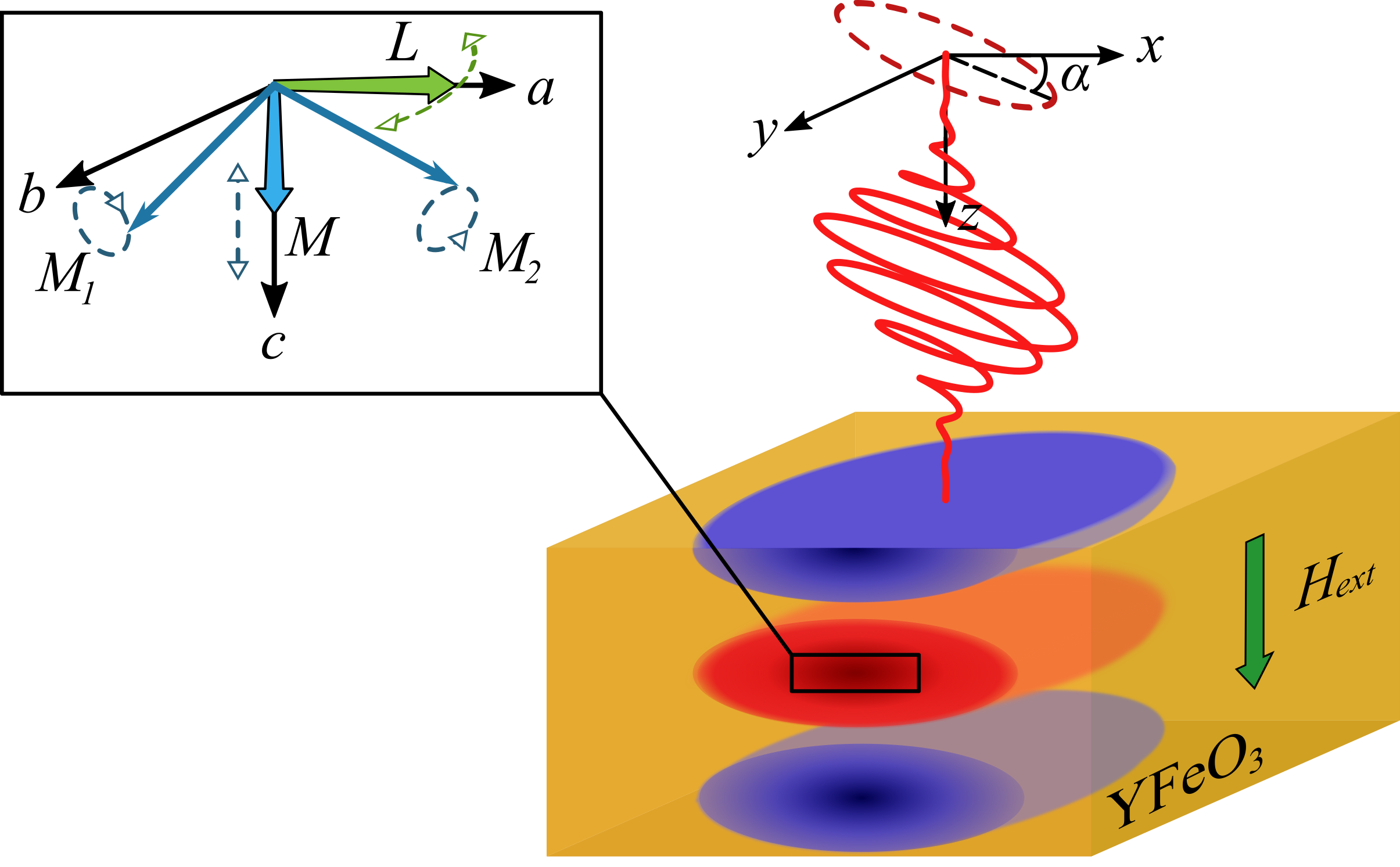}
    \caption{Schematic representation of the configuration under study and the dynamics of antiferromagnetism and magnetization vectors in the quasi-antiferromagnetic mode (inset).}
    \label{axis_config}
\end{figure}

IFE can be described in terms of the effective magnetic field $\mathbf{H}_\mathrm{IFE}$ induced by light:
\begin{equation}
\label{IFE}
\mathbf{H}_\mathrm{IFE}=-\frac{i g}{16 \pi M_0}[\mathbf{E} \times \mathbf{E^*}],
\end{equation}
where $g$ is the magneto-optical gyration, $M_0$ is the saturation magnetization, $\mathbf{E}$ is the electric field of incident light and $\mathbf{E^*}$ is its complex conjugate. Therefore, if the polarization of the electromagnetic field varies inside the magnetic material, $\mathbf{H}_\mathrm{IFE}^0$ becomes also inhomogeneous. Let us consider a biaxial crystal in which the incident light propagates along one of the optical crystal axes. For the definiteness, we will consider yttrium orthoferrite (YFeO$_3$) material with $\boldsymbol{c}$ axis perpendicular to the sample plane (Fig.~\ref{axis_config}). In such configuration the diagonal elements of the permittivity tensor are as follows: $\varepsilon_{xx}=2.365$, $\varepsilon_{yy}=2.4$, $\varepsilon_{zz}=2.337$, $\varepsilon_{xy}=-ig$, and $\varepsilon_{yx}=ig$ (other elements equal zero). YFeO$_3$ belongs to the group of rare-earth orthoferrites with orthorhombic crystal structure. Below the Neel temperature YFeO$_3$ behaves like a weak ferromagnet with two Fe$^{3+}$ sublattices coupled antiferromagnetically by an exchange interaction and aligned along the $\boldsymbol{a}$ crystal axis. The presence of the Dzyaloshinskii–Moriya interaction leads to a slight canting of the neighboring Fe$^{3+}$ spins by an angle of $0.5$ degrees in a way to give a material small macroscopic magnetization along the $\boldsymbol{c}$ crystal axis.

The polarization conversion in the material is described using the Jones matrices~\cite{tabor1970visible,zvezdin1997modern}. It allows one to calculate the distribution of electric field along the coordinate of light propagation knowing the initial polarization state $\mathbf{E}_z = \hat{J} \mathbf{E}_{z=0}$. For the configuration under study $\hat{J}$ has the following form:
\begin{equation}
\label{Jones}
\hat{J}=
\left(
\begin{array}{cc}
    \cos{\frac{\phi}{2}}-i\cos{\tau}\sin{\frac{\phi}{2}} & -\sin{\tau}\sin{\frac{\phi}{2}} \\
    \sin{\tau}\sin{\frac{\phi}{2}} & \cos{\frac{\phi}{2}}+i\cos{\tau}\sin{\frac{\phi}{2}}
\end{array}
\right),
\end{equation}
where $\cos{\tau}=(1-\eta^2)(1+\eta^2)^{-1}$, $\phi=k_0 z (n_+-n_-)$, $k_0$ is wavevector of light in vacuum, $\eta=2g\left(\Delta \varepsilon-\sqrt{\Delta\varepsilon^2+4g^2}\right)^{-1}$, $n_{\pm}^2=\frac{1}{2}\left(\varepsilon_{xx}+\varepsilon_{yy} \pm \sqrt{\Delta\varepsilon^2+4g^2}\right)$, $\Delta\varepsilon=\varepsilon_{xx}-\varepsilon_{yy}$, $z$ is the propagation coordinate of the light. Let us take the initial polarization of the incident light in a form of $(E_x, E_y)_{z=0}=(\cos{\alpha}, \sin{(\alpha)}e^{i\psi})$ where $\alpha$ describes the angle between the polarization of the incident light and $\boldsymbol{a}$ crystal axis and $\psi$ is the ellipticity angle that sets the retardation between two orthogonal polarizations. For example, $\alpha=45^\circ$ and $\psi=0$ correspond to the linear initial polarization at an angle of $45^\circ$ to the $\boldsymbol{a}$ crystal axis, $\alpha=45^\circ$ and $\psi=\pm 90^\circ$ sets the circularly polarized incident light. 

\begin{figure}[h!]
\includegraphics[width=0.5\textwidth]{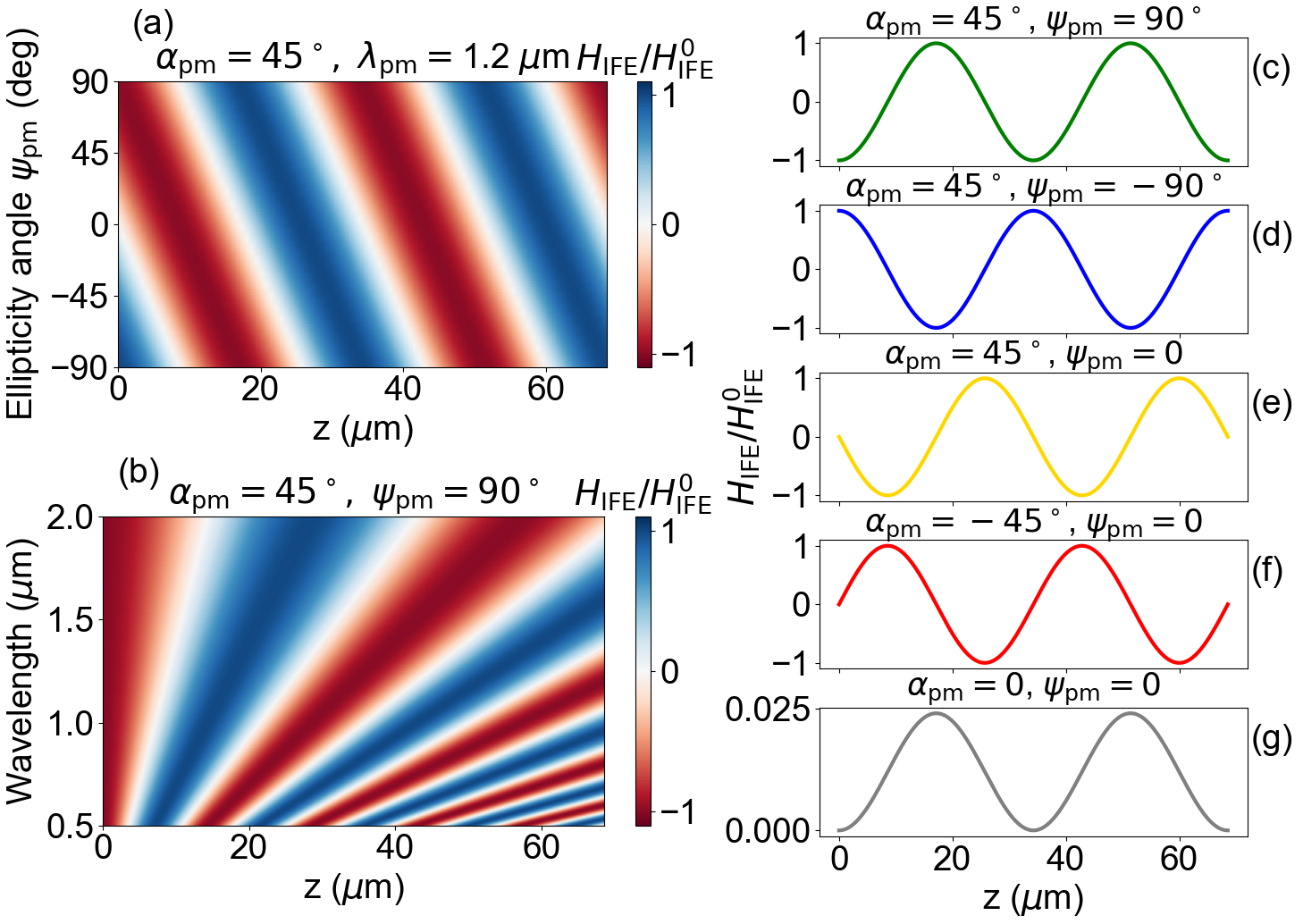}
\caption{IFE-induced effective magnetic field distribution inside the $\boldsymbol{c}$-cut yttrium orthoferrite (normalized at $H_\mathrm{IFE}^{0}$) vs. (a) the ellipticity and (b) the wavelength of the incident light. (c)-(g) figures show effective magnetic field distributions generated by the illumination of (c) right circularly polarized ($\alpha=45^\circ$, $\psi=90^\circ$); (d) left circularly polarized ($\alpha=45^\circ$, $\psi=-90^\circ$); (e) linearly polarized at an angle of $\alpha=45^\circ$ to the $\boldsymbol{a}$ crystal axis ($\psi=0$); (f) linearly polarized at an angle of $\alpha=-45^\circ$ to the $\boldsymbol{a}$ crystal axis ($\psi=0$); (g) linearly polarized along the $\boldsymbol{a}$ crystal axis incident light ($\alpha=0$, $\psi=0$).}
\label{induced}
\end{figure}

Since the investigated medium simultaneously possesses optically anisotropic and gyrotropic properties the resulting distribution of the optical electric field as well as the IFE field depend strongly on the incident light polarization and wavelength (Fig.~\ref{induced}). Using Eq.~\eqref{IFE},\eqref{Jones} one can obtain:
\begin{equation}
\label{hifepsi}
    H_\mathrm{IFE}=H_\mathrm{IFE}^0 \sin{(2\alpha_\mathrm{pm})} \sin{\left(k_\mathrm{pm} \Delta n z+\psi_\mathrm{pm}\right)},
\end{equation}
where $\Delta n=\sqrt{\varepsilon_{xx}}-\sqrt{\varepsilon_{yy}}$ in the approximation $|g|\ll|\varepsilon_{xx}-\varepsilon_{yy}|$ valid for the anisotropic antiferromagnets in a wide spectral range~\cite{zvezdin1997modern}, $H_\mathrm{IFE}^{0}=-g|\mathbf{E}|^2 /(16 \pi M_0)$ is the IFE magnitude for circularly polarized light, $k_\mathrm{pm}$, $\psi_\mathrm{pm}$ and $\alpha_\mathrm{pm}$ are the pump wavevector in vacuum, its ellipticity and polarization angles, correspondingly. Therefore, in this case light induce $H_\mathrm{IFE}$ which oscillates along the crystal plate thickness in accordance to a harmonic law. Initial phase of the spatial distribution equals to the ellipticity angle of the incident light  $\psi_\mathrm{pm}$. Consequently, by varying the ellipticity of the incident light one can modify the initial phase of the induced magnetic field spatial oscillations along the crystal thickness (Fig.~\ref{induced}a,c-f). The second possibility is to tune the spatial frequency of such oscillations by changing the wavelength of the incident light (Fig.~\ref{induced}b). At the same time, amplitude of $H_\mathrm{IFE}$ is determined by the polarization angle of the incident light. 

It is interesting that even for the case in which the incident light is polarized linearly along one of the crystal axis ($\alpha_\mathrm{pm}=\psi_\mathrm{pm}=0$), it is still possible to observe small IFE oscillations (Fig.~\ref{induced}g) due to the mutual action of the Faraday rotation and optical birefringence. However, this effect is rather weak and will not be taken into account further.

Let us now analyze how the inhomogeneous pattern of $H_\mathrm{IFE}$ launches the spin dynamics in the antiferromagnetic crystal. Spin dynamics in an antiferromagnet can be described in terms of a unitary antiferromagnetism vector $\mathbf{L}=(\mathbf{M}_1-\mathbf{M}_2)/(2 M_0)$, where $\mathbf{M}_1$ and $\mathbf{M}_2$ are the magnetic moments of sublattices, and $M_0$ is the saturation magnetization of each sublattice. In the spherical coordinate system with a polar axis aligned along the $\boldsymbol{c}$ crystal axis and an azimuthal axis along the $\boldsymbol{a}$ crystal axis $\mathbf{L}$ has the following components:
\begin{equation}
\label{antiferro2}
    \mathbf{L}=(\sin{\theta} \cos{\varphi}, \sin{\theta} \sin{\varphi}, \cos{\theta}).
\end{equation}

Dynamics of $\mathbf{L}$ can be described in the terms of the Lagrangian $\mathcal{L}$ and Rayleigh dissipation functions $\mathcal{R}$~\cite{zvezdin2020ultrafast,Andreev:1980,zvezdin1979dynamics,kim2017field}:
\begin{equation}
    \label{lagrang}
    \mathcal{L}=\frac{\chi_{\perp}}{2 \gamma^2} \left(\dot{\theta}^2 + (\dot{\varphi}-H_\mathrm{dr})^2 \sin^2{\theta}\right) - \mathcal{U}(\theta, \varphi),
\end{equation}
\begin{equation}
    \mathcal{R}=\frac{\zeta M_0}{2 \gamma} (\dot{\theta}^2 + \dot{\varphi}^2 \sin^2{\theta}),
\end{equation}
where $\chi_{\perp}$ is the transverse susceptibility of the material, $\gamma=1.73 \cdot 10^7$ is the gyromagnetic ratio, $\zeta$ is the dimensionless Gilbert damping constant, and $H_{\mathrm{dr}}$ is the external driving force that appears due to the IFE and for the Gaussian optical pulse is:
\begin{equation}
\label{gaussian}
    H_{\mathrm{dr}} = H_{\mathrm{IFE}}(z) \exp{\left(-\frac{t^2}{2 \Delta t^2}\right)},
\end{equation}
where $\Delta t$ is pulse  duration, and $H_{\mathrm{IFE}}(z)$ is given by Eq.~\eqref{hifepsi}. The first term in Eq.~\eqref{lagrang} describes the kinetic energy of the system while the second one $\mathcal{U}$ is the potential energy which consists of the following components: $\mathcal{U}(\theta, \varphi) = U_a + U_m$, where $U_a$ describes the energy of crystallographic anisotropy, $U_m$ is the Zeeman energy of magnetization interaction with the effective magnetic field. This approach was used for the description of the spatially-homogeneous spin dynamics, however it can also be extended for the considered system where $U_s$ slowly varies in space at the scales of several micrometers. The resulting Euler-Lagrange equations along with the subsequent linearization lead to the following simultaneous equations that describe dynamics of the antiferromagnetism vector:
\begin{equation}
\begin{cases}
    \label{eqsys}
    \ddot{\theta}_1 + \frac{2}{\tau} \dot{\theta}_1 + \omega_2^2 \theta_1 = 0, \\
    \ddot{\varphi}_1 + \frac{2}{\tau} \dot{\varphi}_1 + \omega_1^2 \varphi_1 = \gamma \dot{H}_{\mathrm{dr}},
\end{cases}
\end{equation}
where $\theta_1$ and $\varphi_1$ are the small deviations of angular variables $\theta$ and $\varphi$, $2/\tau = \zeta M_0 \gamma / \chi_\perp$, $\omega_1=\gamma\left(H_\mathrm{d}\left(H_\mathrm{d}-H_\mathrm{ext}\right)+2K_b/\chi_\perp\right)^{\frac{1}{2}}$ and $\omega_2=\gamma\left(H_\mathrm{d}\left(H_\mathrm{ext}-2H_\mathrm{d}\right)+H_\mathrm{ext}^2-2K_{ac}/\chi_\perp\right)^{\frac{1}{2}}$ are frequencies of the quasi-antiferromagnetic and ferromagnetic modes correspondingly, $H_\mathrm{ext}=3$~kOe is the external magnetic field, $H_\mathrm{d}=150$~kOe is the effective magnetic field of Dzyaloshinskii–Moriya interaction, and $K_{ac}$ is the anisotropy constant. Notice that spatial derivatives of $\theta_1$ and $\varphi_1$ are neglected in Eq.~\eqref{eqsys} due to the slow variation in space so that coordinate $z$ appears in Eq.~\eqref{eqsys} only as a parameter describing the driving field (Eq.~\eqref{gaussian}).

Let us focus on the quasi-antiferromagnetic mode. The calculated frequency $\omega_1$ is about 0.52 THz which is in a good agreement with a recently reported experimental values~\cite{zhou2012terahertz,kim2014coherently}. The solution for the second equation of this system is:
\begin{multline}
\label{solution}
    \varphi_1=-\frac{H_\mathrm{IFE}(z)\chi_\perp \gamma\varkappa \sqrt{2\pi}}{\Delta \omega} \exp{\left(-\frac{\omega_1^2-1/\tau^2}{2 \Delta \omega^2}\right)}\times \\ \times\exp{\left(-\frac{t}{\tau}\right)}
    \sin{(\omega_1 t + \beta)},
\end{multline}
where $\Delta \omega=10^{13}$~rad/s is the laser pulse spectral width which corresponds to the pulse duration $\Delta t=100$~fs and $\varkappa=\sqrt{1+1/(\omega_1\tau)^2}$,
\begin{multline}
\label{beta}
\beta=\arcsin\Bigg(\exp\left(-\frac{\tau^2}{\Delta\omega^2}\right)\times \\ \times\sin\left(\frac{\omega_1}{\tau \Delta\omega^2} +\arcsin\varkappa^{-1}\right)\Bigg).
\end{multline} 
The first equation in the system \eqref{eqsys} gives $\theta_1=0$. Thus, the antiferromagnetism vector $\mathbf{L}$ oscillates in $\boldsymbol{ab}$ crystallographic plane (the inset on Fig.~\ref{axis_config}). The dynamic magnetization component is given by the expression: $\mathbf{M_d} =\chi_\perp [\mathbf{L} \times \dot{\mathbf{L}}]/\gamma$. Using the form of antiferromagnetism vector given by Eq.~\eqref{antiferro2} in accordance with the obtained solutions (Eq.~\eqref{solution} and \eqref{beta}) one can derive the components of the dynamic magnetization $\mathbf{M_d} = (0, 0, M_\mathrm{d})$, where $M_\mathrm{d}$ is given by:
\begin{equation}
\label{dynmagnz}
    M_\mathrm{d} = H_\mathrm{A} \chi_\perp \exp{\left(-\frac{t}{\tau}\right)} \sin\left(\omega_1 t+\xi\right),
\end{equation}
where
\begin{equation}
\label{H_a}
    H_\mathrm{A} = \frac{H_\mathrm{IFE}(z)\omega_1\varkappa^2\sqrt{2\pi}}{\Delta\omega}  \exp{\left(-\frac{\omega_1^2-1/\tau^2}{2 \Delta \omega^2}\right)}
\end{equation}
\begin{equation}
\label{xi}
\xi=\beta-\arcsin\varkappa^{-1}
\end{equation}

Thus the amplitude and the initial phase of such quasi-antiferromagnetic mode has a strong spatial dependence so that such modes are labeled as inhomogeneous quasi-antiferromagnetic (iq-AFM) modes hereafter. Full adjustability of the driving force $H_\mathrm{IFE}$ spatial distribution (Fig.~\ref{induced}a,b) allows one to excite iq-AFM modes with harmonic distribution along $\boldsymbol{c}$ axis. 

There is still much uncertainty how optical birefringence of the probe together with a complex mode spatial profile act on the observed probe signal although some studies describing certain special cases numerically were carried out~\cite{de2017effect}. Here we aim to develop a general analytical theory of the pump-probe technique in the presence of birefringence. It is especially important for the spin modes characterized by vanishing integral $\int_{0}^{h} M_\mathrm{d}(z) \,dz = 0$ ($h$ is the crystal thickness) that can not be detected in isotropic materials since the full Faraday rotation of the probe polarization is zero in this case. 

We consider the probe polarization aligned along the $\mathbf{a}$ crystal axis. For the homogeneous magnetization of the optically anisotropic crystal, the Faraday rotation has an oscillatory behavior along the optical axis due to the birefringence~\cite{zvezdin1997modern}:
\begin{equation}
\Phi=-\frac{g}{\Delta\varepsilon}\sin{(k_\mathrm{pb}\Delta n z)},
\end{equation}
where $k_\mathrm{pb}=2 \pi/\lambda_\mathrm{pb}$. For the amplitude of magnetization time precession $M_\mathrm{d}^\mathrm{A}(z)$ oscillating along $z$ axis the gyration acquires an oscillating term as well: $g(z)=g_0M_\mathrm{d}^\mathrm{A}(z)/M_0$  and the resulting Faraday rotation of the probe caused by the oscillating magnetization is calculated by:
\begin{multline}
\label{sumrot}
\Phi_\mathrm{osc}=-\int_0^h \frac{d\Phi}{dz} d z= \\
=-\int_0^h \frac{g_0}{\Delta\varepsilon}\frac{M_\mathrm{d}^\mathrm{A}(z)}{M_0}k_\mathrm{pb}\Delta n \cos{(k_\mathrm{pb}\Delta n z)}dz.
\end{multline}

It should be noted that while the oscillating part of magnetization is induced via IFE in the present case, the consideration below is valid for any mechanism of the harmonic magnetization distribution.

\begin{figure}
\centering
\includegraphics[width=0.9\linewidth]{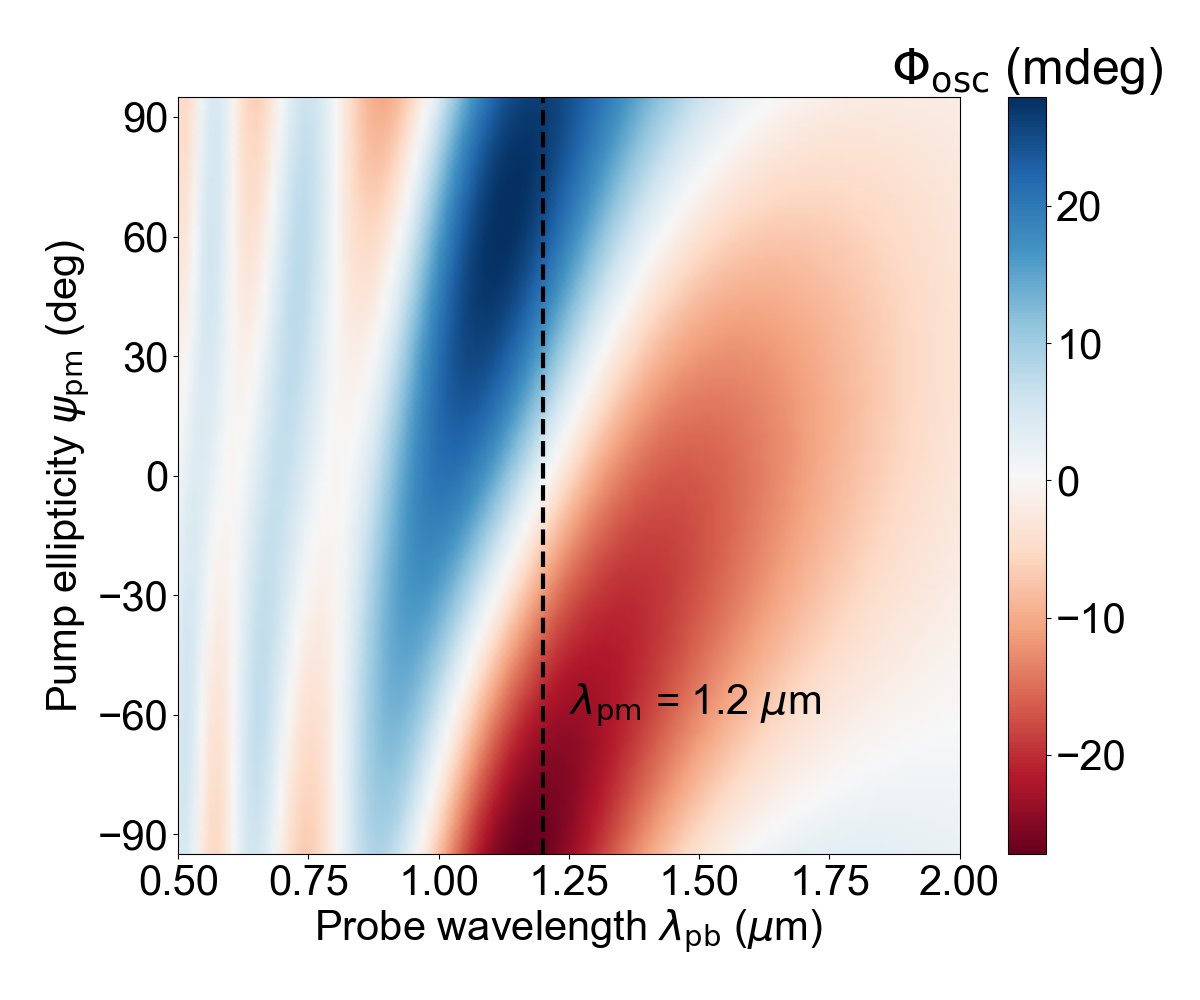}\\
(a)\\
\includegraphics[width=0.9\linewidth]{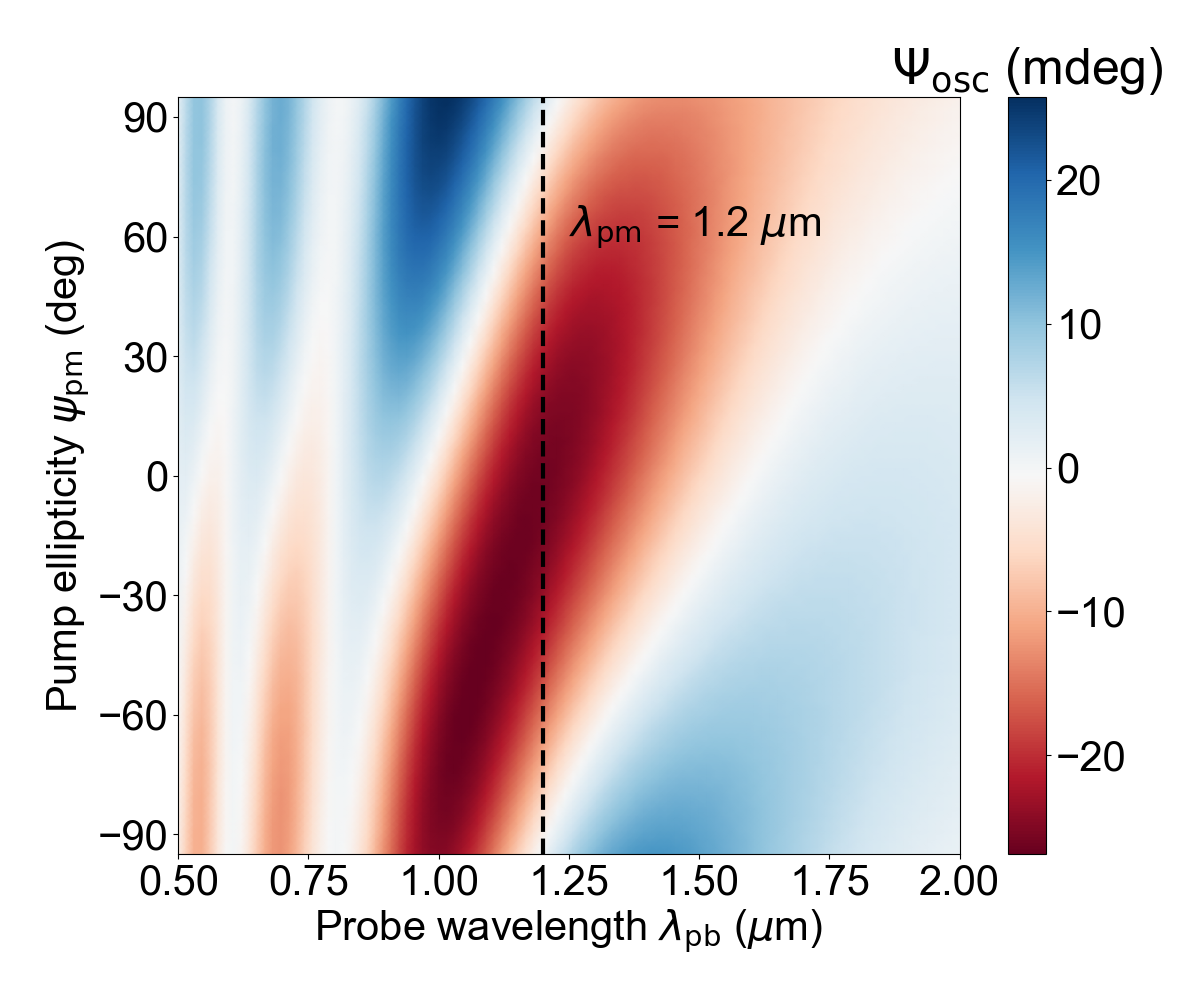}\\
(b)
\caption{Amplitude (with respect to the sign) of the probe polarization $\Phi_\mathrm{osc}$ (a) and induced ellipticity $\Psi_\mathrm{osc}$ (b) temporal oscillations in the dependence on the wavelength $\lambda_\mathrm{pb}$ of the probe and the ellipticity angle $\psi_\mathrm{pm}$ of the pump pulses. The wavelength of pump pulse ($\lambda_\mathrm{pm} = 1.2~\mu$m) and the length of the crystal ($h=68.6~\mu$m) are chosen in such a way to provide the excitation of the iq-AFM mode.}
\label{fbab}
\end{figure}

According to Eqs.~\eqref{hifepsi} and \eqref{dynmagnz} $M_\mathrm{d}^\mathrm{A}(z)\sim \sin{(k_\mathrm{pm}\Delta n z + \psi_\mathrm{pm})}$ and the probe Faraday rotation is determined by the correlation between phases and frequencies of harmonic functions under the integral in Eq.~\eqref{sumrot}. Actually, Eq.~\eqref{sumrot} gives a clear picture of how probe wavelength could be tuned to see the desired iq-AFM mode and explains probing of the spin modes with $\int_{0}^{h} M_\mathrm{d}^\mathrm{A}(z) \,dz = 0$ that is impossible in an isotropic medium regardless of the probe wavelength. Fig.~\ref{fbab}a shows that for every value of pump ellipticity it is possible to tune the wavelength of the probe pulse in order to detect the spin oscillations caused by the nonuniform iq-AFM mode. The value of relation $M_\mathrm{d}^\mathrm{A}/M_0 \sim 0.01$ was chosen in the present consideration as a typical value of magnetization precession amplitude for YFeO$_3$. Notice that in the case of single color pump-probe experiment ($\lambda_\mathrm{pm}=\lambda_\mathrm{pb}=1.2~\mu$m)  (black dashed line in Fig.~\ref{fbab}a) if the initial pump polarization is linear ($\psi_\mathrm{pm}=0$) the amplitude of Faraday effect oscillations is zero due to the presence of two harmonic functions of different symmetry in the integral of Eq.~\eqref{sumrot}. The maximum values of the probe Faraday effect oscillations take place near the condition of coincidence of the effective magnetic field spatial distribution $H_\mathrm{IFE}$ with the derivative of the probe pulse polarization conversion ($\cos{(k_\mathrm{pb}\Delta n z)}$), however, the optimal wavelength is slightly blue-shifted (see Fig.~\ref{fbab}a) due to the $k_\mathrm{pb}$ multiplication under the integral in Eq.~\ref{sumrot}.

The expression for oscillating part of the magnetically induced probe ellipticity can be derived similarly to Eq.~\eqref{sumrot}. 

\begin{figure}[h]
    \centering
    \includegraphics[width=0.49\textwidth]{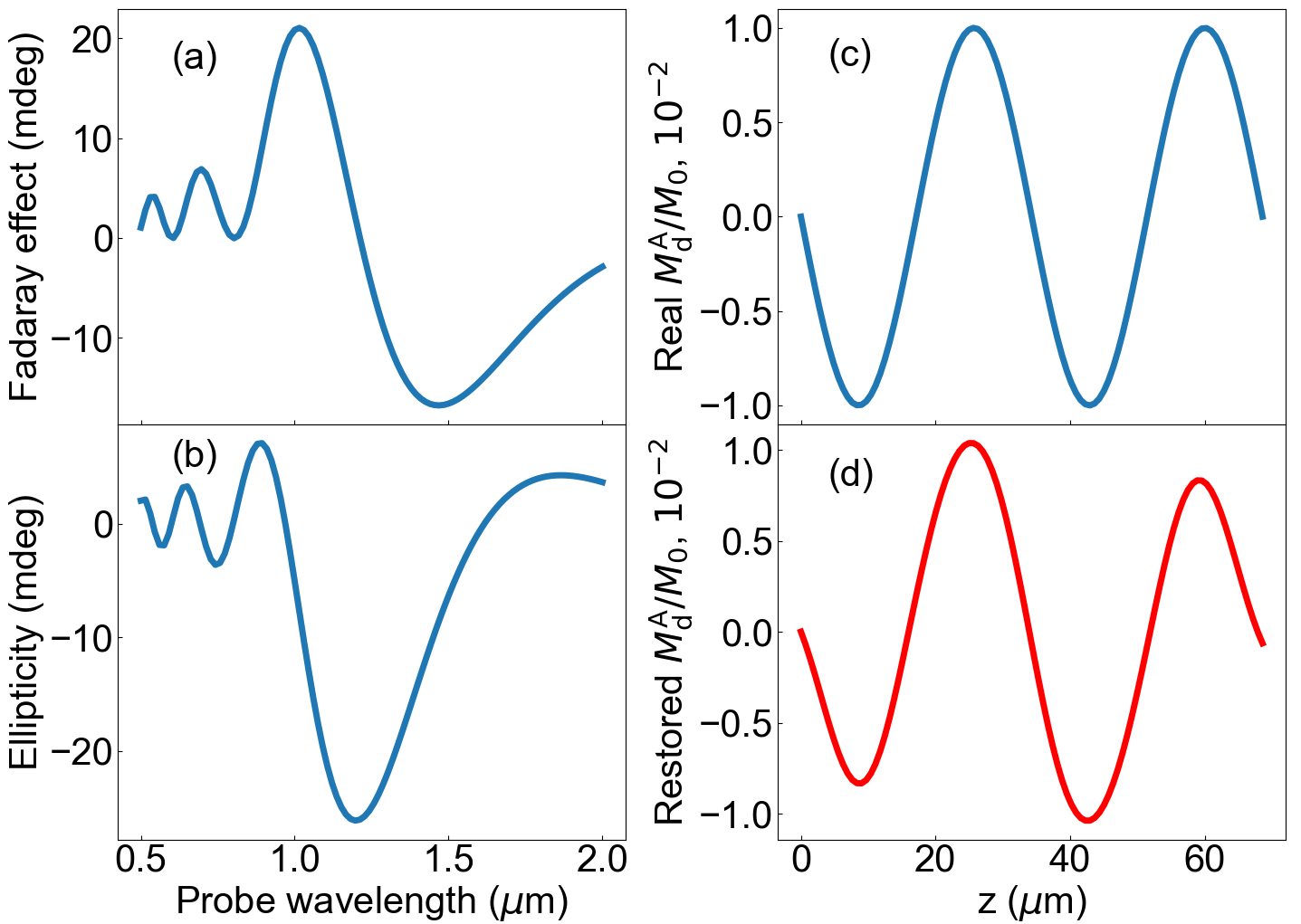}
    \caption{The distribution of the magnetization precession amplitude induced by a pump pulse with a certain wavelength in anisotropic material could be restored (d) using the measured spectral dependencies of the amplitude of polarization plane (a) and ellipticity (b) oscillations of a probe pulse. (c) The real distribution of the magnetization precession amplitude in the medium.}
    \label{recz}
\end{figure}

Finally, the Faraday rotation and ellipticity of the probe can be combined by the following expression: 
\begin{equation}
\label{sumint}
\frac{\Delta\varepsilon\left(\Phi_\mathrm{osc}+i\Psi_\mathrm{osc}\right)}{g_0 k_\mathrm{pb}\Delta n}=-\int_0^h \frac{M_\mathrm{d}^\mathrm{A}(z)}{M_0} \exp{\left(ik_\mathrm{pb}\Delta nz\right)}dz.
\end{equation}

Since $M_\mathrm{d}^\mathrm{A}(z)$ is defined only in the region of existence of an anisotropic material, the integral in Eq.~\eqref{sumint} could be expanded to the infinite limits with $M_\mathrm{d}^\mathrm{A}=0$ for $z<0$ and $z>h$. Therefore, the distribution of the induced via IFE magnetization could be found after the Fourier-transorm of the left part in Eq.~\eqref{sumint}:
\begin{multline}
\label{reconstr}
M_\mathrm{d}^\mathrm{A}(z)=-\frac{M_0}{2\pi}\int_{-\infty}^{+\infty}\frac{\Delta\varepsilon\left(\Phi_\mathrm{osc}+i\Psi_\mathrm{osc}\right)}{g_0 k_\mathrm{pb}\Delta n}\times\\
\times\exp{\left(-ik_\mathrm{pb}\Delta nz\right)}d(k_\mathrm{pb}\Delta n).
\end{multline}

The obtained expression is particularly useful for studying an inhomogeneous magnetization. It allows one to restore the distribution of the inhomogeneous magnetization, including the one induced by the incident pump pulse via IFE $H_\mathrm{IFE}(z)$, using the measured spectral dependencies of the Faraday effect $\Phi_\mathrm{osc}$ and ellipticity $\Psi_\mathrm{osc}$ of the probe pulse. Fig.~\ref{recz} shows the whole process. Notice that even if such measurements are performed in a limited spectral range (Fig.~\ref{recz}(a-b)), the restored through Eq.~\eqref{reconstr} effective magnetic field (Fig.~\ref{recz}(d)) is in a solid agreement with the real one (Fig.~\ref{recz}(c)).

Let us now briefly summarize what unique possibilities for the optical launching and probing of the spin dynamics are provided by the birefringence. 
Optical birefringence leads to modification of the incident pump pulse polarization inside the crystal which is responsible for spatial distribution of the effective magnetic field induced via the inverse Faraday effect that launches magnetization oscillations. The spatial distribution of the IFE inside the crystal can be tuned by variation of the pump wavelength and polarization. 

Importantly, it is possible to investigate the temporal dynamics of such inhomogeneous modes, including the modes characterized with $\int_{0}^{L} M_\mathrm{d}^\mathrm{A}(z) \,dz = 0$ that can not be seen in isotropic materials, in the conventional pump-probe experiments. Optical birefringence affects the magneto-optical Faraday rotation allowing to tune the wavelength of the probe to achieve sensitivity to any of the excited modes. A novel approach is presented for reconstruction of the inhomogeneous magnetization spatial distribution, which can be both static (for example, in the case of domain structure~\cite{pyatakov2011magnetically,logginov2007magnetoelectric}) or dynamic (induced by the pump pulse as in the considered scenario). Even if the spectral region where the probe measurements could be performed is limited, one can achieve great agreement with the real magnetization distribution in the investigated medium.

This work was financially supported by the Ministry of Science and Higher Education of the Russian Federation, Megagrant project N 075-15-2019-1934.

\providecommand{\noopsort}[1]{}\providecommand{\singleletter}[1]{#1}%

\end{document}